\global\def\draftcontrol{0}

   \def\versionno{ n=2 deformed -- draft   }
\catcode`\@=11

\expandafter\ifx\csname draftcontrol\endcsname\relax\global\def\draftcontrol{0}
\fi

{\count255=\time\divide\count255 by 60
\xdef\hourmin{\number\count255}
\multiply\count255 by-60\advance\count255 by\time
\xdef\hourmin{\hourmin:\ifnum\count255<10 0\fi\the\count255}}
\def\draftdate{\number\month/\number\day/\number\year\ \ \ \hourmin }

\newcommand\makepapertitle{\par
  \begingroup
    \renewcommand\thefootnote{\@fnsymbol\c@footnote}%
    \def\@makefnmark{\rlap{\@textsuperscript{\normalfont\@thefnmark}}}%
    \long\def\@makefntext##1{\parindent 1em\noindent
            \hb@xt@1.8em{%
                \hss\@textsuperscript{\normalfont\@thefnmark}}##1}%
     \newpage
     \global\@topnum\z@   % Prevents figures from going at top of page.
     \@makepapertitle
     \thispagestyle{empty}\@thanks
  \endgroup
  \setcounter{footnote}{0}%
  \global\let\thanks\relax
  \global\let\makepapertitle\relax
  \global\let\@makepapertitle\relax
  \global\let\@thanks\@empty
  \global\let\@author\@empty
  \global\let\@date\@empty
  \global\let\@title\@empty
  \global\let\title\relax
  \global\let\author\relax
  \global\let\date\relax
  \global\let\and\relax
  \def\version{\let\version\@version\@gobble}
}
\def\@makepapertitle{%
  \newpage
   \ifnum\draftcontrol=1 {}
   \version\versionno
   \vskip 3em%
   \else
   \hfill\hbox to 3cm {\parbox{4cm}{\@pubnum}\hss}%
   \vskip 3em%
   \fi
   \begin{center}%
   \let \footnote \thanks
     {\LARGE {\@title}}%
     \vskip 1.5em%
     {\normalsize%\large
       \lineskip .5em%
       \begin{tabular}[t]{c}%
         \@author
       \end{tabular}\par}%
     \vskip 1.5em%
     {\@bstract}%
     \end{center}%
     \vskip 1.5em 
     \@date%
   \par
}

\gdef\@pubnum{}
\def\pubnum#1{%
  \gdef\@pubnum{#1}}

\gdef\@bstract{}
\def\Abstract#1{%
  \gdef\@bstract{%
   \parbox{\textwidth-0pc}{%
   \centerline{\bf Abstract}\penalty1000
   \noindent%\abstractfont \baselineskip=12pt
   \renewcommand\baselinestretch{1.0}
   {#1}}}
}

\def\ps@paper{\let\@mkboth\@gobbletwo%
     \ifnum\draftcontrol=1
	\def\@oddfoot{\hbox to \textwidth{\tiny \versionno \hfil\tiny\draftdate}%
	\hskip -\textwidth \hbox to \textwidth{\hfil\rm\thepage\hfil}}%
     \else\def\@oddfoot{\hbox to \textwidth{\hfil\rm\thepage\hfil}}
     \fi
     \let\@evenfoot\@oddfoot
}
\def\body{\clearpage
          \pagestyle{paper}
	}
\def\@version#1{\ifnum\draftcontrol=1
\typeout{}\typeout{#1}\typeout{}
\vskip3mm\centerline{\hbox{\fbox{\normalsize{\tt DRAFT -- #1 -- }
                   {\draftdate}}}}\vskip3mm
\fi}
\let\version\@version
\long\def\eqlabel#1{\ifnum\draftcontrol=1
                    \tag@false  % there are some problems with multline without this
                    \tag*{(\theequation) \hbox to -0.2cm{\hspace{0cm}\small{#1}\hss}}
                    \refstepcounter{equation} 
                    \edef\@currentlabel{\theequation}
                    \ltx@label{#1}          % use old LaTeX \label instead of new definition
                    \else
                    \label{#1}
                    \fi
                    }
\let\st@bibitem\@bibitem
\let\st@lbibitem\@lbibitem
\ifnum\draftcontrol=1
  \def\@bibitem#1{%
    \st@bibitem{#1}\a@@label{#1}\ignorespaces}
  \def\@lbibitem[#1]#2{%
    \st@lbibitem[#1]{#2}\a@@label{#2}\ignorespaces}
  \def\a@@label#1{%
    \gdef\a@lab{\smash{\normalfont\small#1}}
    \ifvmode
      \if@inlabel
        \global\setbox\@labels\hbox{%
          \llap{\a@lab\let\a@lab\relax
                \kern\@totalleftmargin\kern\marginparsep}%
          \box\@labels}%
      \fi
    \fi}
\fi
\documentclass[12pt,letterpaper]{article}
\usepackage{amsmath,amssymb,array,calc,rotating,epsfig,psfrag}
\usepackage[nosort]{cite}
\ifnum\draftcontrol=1
\fi
\renewcommand\baselinestretch{1.25}
\renewcommand\section{\@startsection {section}{1}{\z@}%
                                   {-3.5ex \@plus -1ex \@minus -.2ex}%
                                   {2.3ex \@plus.2ex}%
                                   {\normalfont\large\bfseries}}
\renewcommand\subsection{\@startsection{subsection}{2}{\z@}%
                                   {-3.25ex\@plus -1ex \@minus -.2ex}%
                                   {1.5ex \@plus .2ex}%
                                   {\normalfont\normalsize\bfseries}}
\renewcommand\subsubsection{\@startsection{subsubsection}{3}{\z@}%
                                   {-3.25ex\@plus -1ex \@minus -.2ex}%
                                   {1.5ex \@plus .2ex}%
                                   {\normalfont\normalsize\it}}
\renewcommand\paragraph{\@startsection{paragraph}{4}{\z@}%
                                   {-3.25ex\@plus -1ex \@minus -.2ex}%
                                   {1.5ex \@plus .2ex}%
                                   {\normalfont\normalsize\bf}}

\def\revise#1       {\raisebox{-0em}{\rule{3pt}{1em}}%
                     \marginpar{\raisebox{.5em}{\vrule width3pt\
                     \vrule width0pt height 0pt depth0.5em
                     \hbox to 0cm{\hspace{0cm}{%
                     \parbox[t]{4em}{\raggedright\footnotesize{#1}}}\hss}}}}

\def\cala         {{\cal A}}

\def\calb         {{\cal B}}

\def\calf         {{\cal F}}

\def\caln         {{\cal N}}

\def\calp         {{\cal P}}

\def\del          {\partial}

\def\tr           {\mathop{\rm Tr}}

\def\sqr#1#2{{\vcenter{\vbox{\hrule height.#2pt  
 \hbox{\vrule width.#2pt height#1pt \kern#1pt
 \vrule width.#2pt}\hrule height.#2pt}}}}

\def\tQ{\tilde{Q}}
\def\a{\alpha}
\def\r{\rho}
\def\st{\tilde{\sigma}}

\catcode`\@=12

\begin{document}

%%%
%%%%%% text starts here
%%%%%%%%%

\title{Compactifications of the $\caln=2^*$ flow}

\pubnum{%
MCTP-03-06 \\
hep-th/0302107}
\date{February 2003}

\author{
Alex Buchel\\[0.4cm]
\it Michigan Center for Theoretical Physics \\
\it Randall Laboratory of Physics, The University of Michigan \\
\it Ann Arbor, MI 48109-1120 \\[0.2cm]
}

\Abstract{
In hep-th/0004063 Pilch and Warner (PW) constructed $\caln=2$
supersymmetric RG flow corresponding to the  mass
deformation of the $\caln=4$ $SU(N)$ Yang-Mills theory.
In this paper we present exact  deformations of PW flow 
when the gauge theory 3-space is compactified on $S^3$. 
We consider also the case with the gauge theory 
world-volume being $dS_4$ instead of $R^{3,1}$.
The solution is constructed in five-dimensional gauged supergravity 
and is further uplifted to 10d. 
}

%\enlargethispage{1.5cm}

\makepapertitle

\body

\version\versionno

\section{Introduction}
Probably the most intriguing aspect of the gauge theory/string theory 
duality \cite{m9711} 
(see \cite{a9905} for a review) is the fact that it provides 
a dynamical principle for the nonperturbative definition of string theory
in the asymptotically Anti de Sitter spacetime, where there is no notion of an 
$S$-matrix. The best understood example  of this  duality 
is for the $\caln=4$ $SU(N)$ supersymmetric Yang Mill theory.
Given the original correspondence \cite{m9711}, new examples 
can be constructed  by deforming the gauge theory by 
relevant operators. By now there is an extensive literature on 
such, renormalization group (RG) flow deformations 
\cite{a9905}. In \cite{bt} it was suggested 
that the duality can be extended to cases when one deforms the 
gauge theory space-time. Furthermore, in \cite{bds,blw} it was
suggested that gauge theories on nondynamical de Sitter
backgrounds might be relevant for understanding string theory 
in backgrounds with cosmological horizons.  
Unfortunately it is difficult to use space-time 
deformations of \cite{bt,bds,blw} for developing a detailed 
gauge/string theory duality map. The main problem stems 
from the fact that the examples considered there typically 
involve gauge theory with not well understood ultraviolet
properties. It seems desirable to construct 
nontrivial examples of such deformations for 
``simpler'' gauge theories in the UV.

Probably the simplest  candidate is to consider space-time deformations 
of the massive $\caln=4$ RG flow.  In this paper 
we discuss how to construct such deformations for the 
$\caln=2^*$ RG flow of Pilch and Warner \cite{pw}.

We should emphasize that though we concentrate on the 
flow \cite{pw}, the construction presented here 
can be applied to other RG flows. In particular, 
in Appendix we construct the $S^3$  
deformation of the recent non-supersymmetric 
 ($\caln=0^*$) flow \cite{evans}
in five dimensional gauged supergravity. 
While supergravity flow  \cite{evans} is actually
singular, our deformation is completely smooth. 
We also comment on the physical reason for the singularity    
of  this $\caln=0^*$ flow.

The paper is organized as follows. 
In the next section we review the Pilch-Warner RG flow  
in five dimensions, and discuss it's $S^3$ and $dS_4$ 
deformations. In section 3 we discussed the details of 
the 10d uplift of the deformations. We conclude 
in section 4.

\section{$\caln=2^*$ RG flow and its deformations in five dimensions}

\subsection{The gauge theory story}
In the language of four-dimensional $\caln=1$ supersymmetry, the mass
deformed $\caln=4$ $SU(N)$ Yang-Mills theory ($\caln=2^*$) 
in $R^{3,1}$ consists of a vector
multiplet $V$, an adjoint chiral superfield $\Phi$ related by $\caln=2$
supersymmetry to the gauge field, and two additional adjoint chiral
multiplets $Q$ and $\tilde{Q}$ which form the $\caln=2$ hypermultiplet.  In
addition to the usual gauge-invariant kinetic terms for these fields,
the theory has additional interactions and hypermultiplet mass term
summarized in the superpotential\footnote{The classical K\"{a}hler
potential is normalized $(2/g_{YM}^2)\tr[\bar{\Phi}\Phi+
\bar{Q}Q+\bar{\tQ}\tQ]$.}
\begin{equation}
W=\frac{2\sqrt{2}}{g_{YM}^2}\tr([Q,\tQ]\Phi)
+\frac{m} {g_{YM}^2}(\tr Q^2+\tr\tQ^2)\,.
\eqlabel{sp}
\end{equation}
When  $m=0$  the gauge theory is superconformal with $g_{YM}$
characterizing an exactly marginal deformation. The theory has 
classical $3(N-1)$  complex dimensional moduli space. This moduli  
space is protected by supersymmetry against (non)-perturbative quantum 
corrections. With $m\ne 0$, the $\caln=4$ supersymmetry is 
softly broken to $\caln=2$. This mass deformation lifts 
$\{Q,\ \tQ\}$ hypermultiplet moduli directions, leaving the 
$(N-1)$ complex dimensional Coulomb branch of the $\caln=2$ 
$SU(N)$ Yang-Mill theory, 
parameterized by
expectation values of the adjoint scalar
\begin{equation}
\Phi={\rm diag} (a_1,a_2,\cdots,a_N)\,,\quad \sum_i a_i=0\,,
\eqlabel{adsc}
\end{equation}
in the Cartan subalgebra of the gauge group.  For generic values of
the moduli $a_i$ the gauge symmetry is broken to that of the Cartan
subalgebra $U(1)^{N-1}$, up to the permutation of individual $U(1)$
factors. Additionally, the superpotential 
\eqref{sp} induces the RG flow of the gauge coupling.   
While from the gauge theory perspective it is straightforward 
to study this  $\caln=2^{*}$ gauge theory at any point on the 
Coulomb branch \cite{phil}, the PW supergravity flow \cite{pw}
corresponds to a particular Coulomb branch vacuum. 
More specifically, matching the probe computation 
in gauge theory and the dual PW supergravity flow it was argued 
in \cite{bpp} that the appropriate Coulomb branch vacuum  
corresponds to a  linear distribution of the vevs \eqref{adsc}
as
\begin{equation}
a_i\in [-a_0,a_0],\qquad a_0^2=\frac{m^2 g_{YM}^2 N}{\pi}\,, 
\eqlabel{inter}
\end{equation}
with (continuous in the large $N$ limit) 
linear number density       
\begin{equation}
\rho(a)=\frac{2}{m^2 g_{YM}^2}\sqrt{a_0^2-a^2},\qquad 
\int_{-a_0}^{a_0}da \rho(a)=N\,.
\eqlabel{rho}
\end{equation}
Unfortunately, the extension of the $N=2^*$ gauge/gravity
correspondence of \cite{pw,bpp,cl} for vacua 
other than \eqref{rho} is not known.
 
In \cite{bpp,cl} the dynamics of the gauge theory 
on the D3 brane probe in the PW background was studied in details.
It was shown in \cite{bpp} that the probe has one complex dimensional 
moduli space, with bulk induced  metric precisely equal to the 
metric on the appropriate one complex dimensional submanifold of the 
$SU(N+1)$ $\caln=2^*$ Donagi-Witten theory Coulomb branch. 
This one dimensional submanifold 
is parameterized by the expectation value $u$ of the $U(1)$ complex scalar
on the Coulomb branch of the theory where $SU(N+1)\rightarrow U(1)\times 
SU(N)_{PW}$, and the $_{PW}$ subscript denotes that the $SU(N)$ factor is 
in the Pilch-Warner vacuum \eqref{rho}. As $u$ coincides with 
any of the $a_i$ of the PW vacuum, the moduli space metric diverges,
signaling the appearance of the additional massless states. 
Identical divergence is observed \cite{bpp,cl} for the  probe D3-brane  
at the {\it enhancon} singularity of the PW background.
Away from the singularity locus, $u=a\in [-a_0,a_0]$, the gauge theory 
computation of the probe moduli space metric is 1-loop exact. 
This is due to the  suppression of instanton corrections  in the large $N$ 
limit \cite{bpp,b} of $N=2$ gauge theories. 

Consider now the  $R^{3,1}\to R\times S^3$ or $R^{3,1}\to dS_4$ deformations 
of the $N=2^*$ gauge theory. Both deformations introduce a new 
scale, let's call it $\mu$, to the model 
--- the $S^3$ scale in the former case 
and the Hubble parameter in the latter. Depending on the ratio
$\frac \mu m$ we expect an interesting interplay between the 
strongly coupled $N=2^*$ IR dynamics and  the IR curvature 
induced cutoff. For one reason, we expect that for the sufficiently 
high $\mu$ the number density distribution $\rho(a)$ should be 
just a $\delta$-function at zero. In what follows we 
present and indication for this phase transition while 
postponing the detailed analysis for the future.

\subsection{PW RG flow}
The gauge theory RG flow induced by the superpotential \eqref{sp}
corresponds to five dimensional gauged SUGRA flow induced 
by scalars $\a\equiv \ln \r$ and $\chi$. 
The effective 5d action is 
\begin{equation}
S=\int d\xi^5 \sqrt{-g}\left(\frac 14  R-3(\del\a)^2-(\del\chi)^2-
\calp\right)\,,
\eqlabel{action5}
\end{equation} 
where the potential $\calp$ is\footnote{We set the 5d gauged SUGRA coupling 
to one. This corresponds to setting $S^5$ radius $L=2$.} 
\begin{equation}
\calp=\frac{1}{48} \left(\frac{\del W}{\del \a}\right)^2+
\frac{1}{16} \left(\frac{\del W}{\del \chi}\right)^2-\frac 13 W^2\,,
\eqlabel{pp}
\end{equation}
with the superpotential
\begin{equation}
W=-\frac{1}{\r^2}-\frac 12 \r^4 \cosh(2\chi)\,.
\eqlabel{supp}
\end{equation}
The PW geometry \cite{pw} has the flow metric 
\begin{equation}
ds_5^2=e^{2 A} \left(-dt^2 +d\bar{x}^2\right)+dr^2\,.
\eqlabel{flowmetric}
\end{equation}
The scalar equations of motion and the Einstein equations 
can be reduced to the first order equations
\begin{equation}
\begin{split}
\frac {d\a}{d r}&=\frac{1}{12} \frac{\del W}{\del \a}\,,\\
\frac {d\chi}{d r}&=\frac{1}{4} \frac{\del W}{\del \chi}\,,\\
\frac {d A}{d r}&=-\frac 13 W\,.
\end{split}
\eqlabel{pwo}
\end{equation} 

\subsubsection{Asymptotics of the PW flow}
Given the explicit solution of the flow equations \eqref{pwo}
in \cite{pw} is it easy to extract the UV/IR asymptotics.
In the ultraviolet, $r\to +\infty$, we find
\begin{equation}
{\rm UV:}\qquad\qquad \rho\to 1_-,\qquad \chi\to 0_+,\qquad A\to \frac 12\ 
r\,, 
\eqlabel{pwasi}
\end{equation} 
while in the infrared, $r\to 0$
\begin{equation}
{\rm IR:}\qquad\qquad \rho\to 0_+,\qquad \chi\to +\infty,\qquad
A\to -\frac 83 \chi\,.
\eqlabel{pwas0}
\end{equation}

\subsection{Deformations of the PW flow}
Unlike the PW flow, the deformed flows break the supersymmetry and 
are given by second order equations.
From \eqref{action5} we have Einstein equations
\begin{equation}
\frac 14 R_{\mu\nu}=3\del_\mu \a \del_\nu\a+\del_\mu\chi\del_\nu\chi
+\frac 
13 g_{\mu\nu} \calp\,,
\eqlabel{ee}
\end{equation} 
plus the scalar equations
\begin{equation}
\begin{split}
0&=\frac {6}{\sqrt{-g}}\del_\mu \left(g^{\mu\nu} \sqrt{-g}\ \del_\mu\a \right)
-\frac{\del\calp}{\del \a}\,,\\
0&=\frac {2}{\sqrt{-g}}\del_\mu \left(g^{\mu\nu} \sqrt{-g}\ \del_\mu\chi\right)
-\frac{\del\calp}{\del \chi}\,.\\
\end{split}
\eqlabel{scalar}
\end{equation}

We consider two deformations of the flow metric \eqref{flowmetric}:
\begin{equation}
\begin{split}
(a):\qquad ds_5^2&=e^{2 A} \left(-dt^2 +e^{2 B}\ dS_3^2\right)+dr^2\,,\\
\\
(b):\qquad ds_5^2&=e^{2 A} \left(-dt^2 +\cosh^2 t\  dS_3^2\right)+dr^2\,.
\end{split}
\eqlabel{ab}
\end{equation}

In the first case from \eqref{ee},\eqref{scalar} we find
\begin{equation}
\begin{split}
0&=\a''+\left(4 A' +3 B'\right)\a' -\frac 16 \frac{\del\calp}{\del\a}\,,\\
0&=\chi''+\left(4 A' +3 B'\right)\chi' -\frac 12 \frac{\del\calp}{\del\chi}
\,,\\
0&=B''+4 A' B'+3\left(B'\right)^2-2 e^{-2 A -2 B}\,,\\
&\frac 14 A''+\left(A'\right)^2+\frac 34 A' B'=-\frac 13 \calp\,,\\
&- A''-\left(A'\right)^2-\frac 32 A' B'-\frac 34 B''-\frac 34 
\left(B'\right)^2=3\left(\a'\right)^2+\left(\chi'\right)^2 +\frac 13 
\calp\,,
\end{split}
\eqlabel{aeq}
\end{equation}
while in case (b) we find
\begin{equation}
\begin{split}
0&=\a''+4 A'\a' -\frac 16 \frac{\del\calp}{\del\a}\,,\\
0&=\chi''+4 A'\chi' -\frac 12 \frac{\del\calp}{\del\chi}\,,\\
&\frac 14 A''+\left(A'\right)^2-\frac 34 e^{-2 A}=-\frac 13 \calp\,,\\
&- A''-\left(A'\right)^2
=3\left(\a'\right)^2+\left(\chi'\right)^2 +\frac 13 \calp\,.
\end{split}
\eqlabel{beq}
\end{equation}
It is easy to check that above equations are consistent. 
Thus for the deformed flows we could use the same 
scalars as in the PW case.

\subsubsection{Asymptotics of the $S^3$ deformation}
The flow equations are given by \eqref{aeq}. 
The nonsingular in the IR flows are represented by  a  two parameter 
$\{\r_0>0,\chi_0\}$
Taylor series expansion\footnote{Without loss of generality 
we set $A|_{r=0}=0$. This corresponds to rescaling the time coordinate
in \eqref{ab}.} 
\begin{equation}
\begin{split}
e^A&=1+\left(\sum_{i=1}^{\infty}\ \a_i\ r^{2 i}\right)\,,\\
e^B&=r\left(1+\sum_{i=1}^{\infty}\ b_i\ r^{2 i} \right)\,,\\
\r&=\r_0+\left(\sum_{i=1}^{\infty}\ \r_i\ r^{2 i}\right)\,,\\
\chi&=\chi_0+\left(\sum_{i=1}^{\infty}\ \chi_i\ r^{2 i}\right)\,,
\end{split}
\eqlabel{ds0asa}
\end{equation}
with the first terms being 
\begin{equation}
\begin{split}
a_1&=\frac{1}{24}\ \r_0^{-4}+\frac{1}{12}\ \r_0^2\ \cosh(2\chi_0)-
\frac{1}{96}\ \r_0^8\ \sinh^2(2\chi_0)\,,\\
b_1&=-\frac{1}{36}\ \r_0^{-4}-\frac{1}{18}\ \r_0^2\ \cosh(2\chi_0)+
\frac{1}{144}\ \r_0^8\ \sinh^2(2\chi_0)\,,\\
\r_1&=\frac{1}{48}\ \r_0^{-3}-\frac{1}{48}\ \r_0^3\ \cosh(2\chi_0)+
\frac{1}{96}\ \r_0^9\ \sinh^2(2\chi_0)\,,\\
\chi_1&=-\frac{1}{16}\ \r_0^{2} \sinh(2\chi_0)
    +\frac{1}{128}\ \r_0^8\ \sinh(4\chi_0)\,.
\end{split}
\eqlabel{1dsa}
\end{equation}
We expect that for an appropriate choice of $\{\r_0,\chi_0\}$ 
we recover the UV asymptotics \eqref{pwasi}. 
It is tempting to identify the 2 dimensionless 
parameters of the regular in the IR 
flow with the ratio of $m/\mu$ of the gauge theory (the $\chi_0$ parameter), 
and  the $\rho_0$ parameter as a characteristic of  the brane 
distribution (similar to the 
enhancon scale $a_0$ in \eqref{rho}) 
in the IR. Notice, that unlike PW flow, where 
$\chi\to +\infty$ in the IR, here it is consistent to 
choose\footnote{We would like to interpret $\chi_0=0$ flow 
as a supergravity dual to the $\caln=2^*$ 
flow induced by the $\caln=4$ scalar expectation values. 
Typically, scalar expectation value does not give rise to an 
RG flow. Since these scalars are conformal (and thus couple to the 
$S^3$ curvature), given them an expectation value would induce a flow. 
} $\chi_0=0$. In fact $\chi(r)\equiv 0$ is a solution 
to \eqref{aeq}\footnote{ Also, $\chi(r)\equiv 0$ and $\rho(r)\equiv 1$ 
is a trivial solution corresponding to the global $AdS_5$.}.

The nonsingular flows that asymptote
to \eqref{pwasi} would have a well defined (finite) mass,
being a function of $\{\r_0,\chi_0\}$, characterizing  
phases of the model\footnote{The details of the phase structure will 
be discussed elsewhere.}.

\subsubsection{Asymptotics of the $dS_4$ deformation}
The flow equations are given by \eqref{beq}. 
The nonsingular in the IR flows are represented by  a  two parameter 
$\{\r_0>0,\chi_0\}$
Taylor series expansion 
\begin{equation}
\begin{split}
e^A&=r\left(1+\sum_{i=1}^{\infty}\ a_i\ r^{2 i} \right)\,,\\
\r&=\r_0+\left(\sum_{i=1}^{\infty}\ \r_i\ r^{2 i}\right)\,,\\
\chi&=\chi_0+\left(\sum_{i=1}^{\infty}\ \chi_i\ r^{2 i}\right)\,,
\end{split}
\eqlabel{ds0as}
\end{equation}
with the first terms being 
\begin{equation}
\begin{split}
a_1&=\frac{1}{72}\ \r_0^{-4}+\frac{1}{36}\ \r_0^2\ \cosh(2\chi_0)-
\frac{1}{288}\ \r_0^8\ \sinh^2(2\chi_0)\,,\\
\r_1&=\frac{1}{60}\ \r_0^{-3}-\frac{1}{60}\ \r_0^3\ \cosh(2\chi_0)+
\frac{1}{120}\ \r_0^9\ \sinh^2(2\chi_0)\,,\\
\chi_1&=-\frac{1}{20}\ \r_0^{2} \sinh(2\chi_0)
    +\frac{1}{160}\ \r_0^8\ \sinh(4\chi_0)\,.
\end{split}
\eqlabel{1ds}
\end{equation}
As in the case of the $S^3$ deformation it is also consistent 
here to choose $\chi(r)\equiv 0$.

\section{The ten-dimensional solutions}

\subsection{Type IIB SUGRA equations of motion}

We use mostly positive convention for the signature 
$(-+\cdots +)$ and $\epsilon_{1\cdots10}=+1$.
The type IIB equations consist of \cite{schwarz83}:

\noindent $\bullet$\quad  The Einstein equations:
\begin{equation}
R_{MN}=T^{(1)}_{MN}+T^{(3)}_{MN}+T^{(5)}_{MN}\,,
\eqlabel{tenein}
\end{equation}
where the energy
momentum tensors of the dilaton/axion field, $\calb$, the three index
antisymmetric tensor field, $F_{(3)}$, and the self-dual five-index
tensor field, $F_{(5)}$, are given by
\begin{equation}
T^{(1)}_{MN}= P_MP_N{}^*+P_NP_M{}^*\,,
\eqlabel{enmomP}
\end{equation}
\begin{equation}
T^{(3)}_{MN}=
       \frac 18(G^{PQ}{}_MG^*_{PQN}+G^{*PQ}{}_MG_{PQN}-
        \frac 16 g_{MN} G^{PQR}G^*_{PQR})\,,
\eqlabel{enmomG}
\end{equation}
\begin{equation}
T^{(5)}_{MN}= \frac 16 F^{PQRS}{}_MF_{PQRSN}\,.
\eqlabel{enmomF}
\end{equation}

In the unitary gauge $\calb$ is a complex scalar
field and
\begin{equation}
P_M= f^2\partial_M \calb\,,\qquad Q_M= f^2\,{\rm Im}\,(
\calb\partial_M\calb^*)\,,
\eqlabel{defofPQ}
\end{equation}
with
\begin{equation}
f= \frac{1}{ (1-\calb \calb^*)^{1/2}}\,,
\eqlabel{defoff}
\end{equation}
while the antisymmetric tensor field $G_{(3)}$ is given by
\begin{equation}
G_{(3)}= f(F_{(3)}-\calb F_{(3)}^*)\,.
\eqlabel{defofG}
\end{equation}

\noindent 
$\bullet$\quad The Maxwell equations:
\begin{equation}
(\nabla^P-i Q^P) G_{MNP}= P^P G^*_{MNP}-\frac 23\,i\,F_{MNPQR}
G^{PQR}\,.
\eqlabel{tenmaxwell}
\end{equation}

\noindent
$\bullet$\quad The dilaton equation:
\begin{equation}
(\nabla^M -2 i Q^M) P_M= -\frac{1}{ 24} G^{PQR}G_{PQR}\,.
\eqlabel{tengsq}
\end{equation}
\noindent
$\bullet$\quad The self-dual equation:
\begin{equation}
F_{(5)}= \star F_{(5)}\,.
\eqlabel{tenself}
\end{equation}

In addition, $F_{(3)}$ and $F_{(5)}$ satisfy Bianchi identities which
follow from the definition of those field strengths in terms of their
potentials:
\begin{equation}
\begin{split}
F_{(3)}&= dA_{(2)}\,,\\
F_{(5)}&= dA_{(4)}-{\frac 18}\,{\rm Im}( A_{(2)}\wedge
F_{(3)}^*)\,.
\end{split}
\eqlabel{defpotth}
\end{equation}

For the 10d uplift of the RG flows in the 5d gauged SUGRA 
the metric ansatz  and the dilaton is 
basically determined by group theoretical properties 
of the $d=5$ $\caln=8$ scalars, and thus must be the same for  
both the deformed and original PW flows.
Specifically, we assume \cite{pw} that the 10d Einstein frame metric is 
\begin{equation}
\begin{split}
ds_{10}^2&=\Omega^2 ds_5^2 + 4 \frac {(c X_1 X_2)^{1/4} }{\r^3}\biggl(
c^{-1} d\theta^2+\r^6\cos^2\theta \left(\frac {\sigma_1^2}{c X_2}
+\frac{\sigma_2^2+\sigma_3^2}{X_1}\right)+\sin^2\theta\frac {d\phi^2}{X_2}
\biggr)\,,
\end{split}
\eqlabel{10m}
\end{equation}  
where $ds_5^2$ is either the original PW flow metric 
\eqref{flowmetric} or its deformations \eqref{ab}, 
$c\equiv \cosh (2\chi)$. The warp factor is given by
\begin{equation}
\Omega^2=\frac {(c X_1 X_2)^{1/4} }{\r}\,,
\eqlabel{om5}
\end{equation}
and the two functions $X_i$ are defined by
\begin{equation}
\begin{split}
X_1(r,\theta)&=\cos^2\theta+\r(r)^6\cosh(2\chi(r))\sin^2\theta\,,\\
X_2(r,\theta)&=\cosh(2\chi(r))\cos^2\theta+\r(r)^6\sin^2\theta\,.
\end{split}
\eqlabel{x1x2}
\end{equation}
As usual, $\sigma_i$ are the $SU(2)$ left-invariant forms normalized so that 
$d\sigma_i=2 \sigma_j\wedge \sigma_k$.
For the dilaton/axion we have 
\begin{equation}
f=\frac 12 \left(\left(\frac{c X_1 }{X_2}\right)^{1/4}+
\left(\frac{c X_1 }{X_2}\right)^{-1/4}\right),\qquad 
f\calb =\frac 12 \left(\left(\frac{c X_1 }{X_2}\right)^{1/4}-
\left(\frac{c X_1 }{X_2}\right)^{-1/4}\right) e^{2i \phi}\,.
\eqlabel{dilax}
\end{equation} 

The consistent truncation ansatz does not specify the 
(3-) 5-form fluxes. As in \cite{pw} we assume the most general 
ansatz allowed by the global symmetries of the background
\begin{equation}
A_{(2)}=e^{i\phi}\left(a_1(r,\theta)\ d\theta\wedge \sigma_1+a_2(r,\theta)\
 \sigma_2
\wedge \sigma_3+a_3(r,\theta)\ \sigma_1\wedge d\phi+a_4(r,\theta)\ 
d\theta\wedge d\phi\right)\,,
\eqlabel{a2}  
\end{equation}
where $a_i(r,\theta)$ are arbitrary complex functions. 
For the 5-form flux we assume 
\begin{equation}
\begin{split}
&(a):\qquad F_5=\calf+\star\calf,\qquad 
\calf=dt\wedge {\rm vol}_{S^3}\wedge d\omega\,,
\\
&(b):\qquad F_5=\calf+\star\calf,\qquad 
\calf=\cosh^3 t\ dt\wedge {\rm vol}_{S^3}\wedge d\omega\,,
\end{split}
\eqlabel{5form}
\end{equation}
where $\omega(r,\theta)$ is an arbitrary function.

We will do all the computation in the natural 
orthonormal frame given by 
\begin{equation}
\begin{split}
&e^1\propto dt,\quad e^2\propto dr,\quad e^3\propto \st_1,
\quad e^4\propto \st_2,\quad e^5\propto \st_3,\\
&e^6\propto d\theta,\quad e^7\propto \sigma_1,
\quad e^8\propto \sigma_2,\quad e^9\propto \sigma_3,\quad
e^{10}\propto  d\phi\,,
\end{split}
\eqlabel{frame}
\end{equation}
where $\st_i$ are again $SU(2)$ left-invariant one forms, such that 
the round $S^3$ metric of unit radius is $(dS^3)^2=\sum\st_i^2$.

As in the PW case, examination of the Einstein equations 
reveals that 2-form potential functions $a_i$ have the following 
properties: $a_4\equiv 0$,  $a_1,a_2$ are pure imaginary, and $a_3$
is real.

\subsection{Lift of $S^3$ deformation}
Explicitly computing Ricci tensor with above ansatz,
we find  nonvanishing components $R_{11}, R_{22}, 
R_{33}=R_{44}=R_{55}, R_{66}, R_{77}, R_{88}=R_{99}, 
R_{1010}, R_{26}=R_{62}$. Given the 5d flow equations  
\eqref{aeq}, we find relations 
\begin{equation}
\begin{split}
&R_{77}+R_{88}=2 R_{11}\,,\\
&R_{11}+R_{33}=0\,.
\end{split}
\eqlabel{ricnon}
\end{equation} 
The 3-form energy-momentum tensor has nontrivial components 
$T^{(3)}_{11}=-T^{(3)}_{33}=-T^{(3)}_{44}=-T^{(3)}_{55},
 T^{(3)}_{22},T^{(3)}_{66}, T^{(3)}_{77}, T^{(3)}_{88}=T^{(3)}_{99}, 
T^{(3)}_{1010}, T^{(3)}_{26}=T^{(3)}_{62} $. The nonvanishing 
components of the dilaton/axion energy-momentum tensor are 
$T^{(1)}_{22},T^{(1)}_{66},T^{(1)}_{1010},T^{(1)}_{26}= T^{(1)}_{62}$. 
Finally, the 5-form energy-momentum  tensor has nonvanishing components 
\begin{equation} 
\begin{split}
&T^{(5)}_{11}=-T^{(5)}_{33}=-T^{(5)}_{44}=-T^{(5)}_{55}=T^{(5)}_{77}
=T^{(5)}_{88}=T^{(5)}_{99}=T^{(5)}_{1010}=\cala_1^2+\cala_2^2\,,\\
&T^{(5)}_{22}=-T^{(5)}_{66}=\cala_2^2-\cala_1^2\,,\\
&T^{(5)}_{26}= T^{(5)}_{62}=2\cala_1 \cala_2\,,  
\end{split}
\eqlabel{5fst}
\end{equation}
where
\begin{equation}
\cala_1\propto \frac{\del\omega}{\del r},\qquad 
\cala_2\propto \frac{\del\omega}{\del \theta}\,.
\eqlabel{aidef}
\end{equation}
Besides Einstein equations, we have nontrivial 5-form Bianchi identity,
dilaton/axion equation \eqref{tengsq}, and 4 equations from the 
Maxwell equation \eqref{tenmaxwell} for components $\{MN\}=\{27,67,710,89\}$.
 
As in \cite{pw} we find the following consistency checks 
on the metric and dilaton/axion ansatz\footnote{
There is a typo in the second equation in \eqref{consch} in \cite{pw}
(eq.(4.3)).}:
\begin{equation}
\begin{split}
&T^{(3)}_{1010}-T^{(3)}_{11}=\frac{e^{-2 i\phi}}{24} G_{MNP}G^{MNP}\,,\\
&R_{1010}-R_{11}=2 |P_{10}|^2-e^{-2 i\phi} \left(\nabla^M-2 i Q^M\right)P_M\,.
\end{split}
\eqlabel{consch}
\end{equation}
Next combination is 
\begin{equation}
R_{1010}-R_{77}-2 |P_{10}|^2=T_{1010}^{(3)}-T^{(3)}_{77}\,.
\eqlabel{r107}
\end{equation}
As in \cite{pw}, we find that \eqref{r107} (and the linearized 
solution of all equations in the UV) is satisfied provided\footnote{
Note that there is a sign typo for $a_3$ in the corresponding 
equations in \cite{pw}, (eq.(4.8)).}
\begin{equation}
\begin{split}
a_1&=- i\ 4\ \tanh(2\chi) \cos\theta\,, \\
a_2&=i\ 4\ \frac{\r^6\sinh(2\chi)}{X_1}\ \sin\theta\cos^2\theta\,,\\
a_3&=-4\  \frac{\sinh(2\chi)}{X_2}\ \sin\theta\cos^2\theta\,.
\end{split}
\eqlabel{aaa}
\end{equation}
Finally, from the  $\{MN\}=\{11,22\}$ Einstein equations 
we find 
\begin{equation}
\begin{split}
&\frac{\del \omega}{\del \theta}=-\frac 32 e^{4 A +3 B} \left(\ln \r\right)'\ 
\sin 2\theta\,,\\
&\frac{\del \omega}{\del r}=\frac 18 e^{4 A +3 B}\ 
\frac{1}{\r^4}\ \biggl(-\r^{12}\sinh^2(2\chi)\sin^2\theta +2\r^6 
\cosh(2\chi)(1+\sin^2\theta)+2\cos^2\theta\biggr)\,. 
\end{split}
\eqlabel{wres}
\end{equation}

We explicitly verified that supplementing the 
metric and  the dilaton/axion ansatz 
of the previous section with \eqref{aaa}, \eqref{wres},
and the 5d flow equations  \eqref{aeq}, all the equations of
10d type IIB supergravity are satisfied.
 
\subsection{Lift of $dS_4$ deformation}
In this case the analysis are similar to those in the 
previous section. Thus we present only the results.
First, we find the same complex functions $a_i$, specifying 
the 2-form potential \eqref{a2}
\begin{equation}
\begin{split}
a_1&=- i\ 4\ \tanh(2\chi) \cos\theta\,, \\
a_2&=i\ 4\ \frac{\r^6\sinh(2\chi)}{X_1}\ \sin\theta\cos^2\theta\,,\\
a_3&=-4\  \frac{\sinh(2\chi)}{X_2}\ \sin\theta\cos^2\theta\,.
\end{split}
\eqlabel{aaa1}
\end{equation}
Second, the $\omega$ in the 5-form potential \eqref{5form} is
  \begin{equation}
\begin{split}
&\frac{\del \omega}{\del \theta}=-\frac 32 e^{4 A } \left(\ln \r\right)'\ 
\sin 2\theta\,,\\
&\frac{\del \omega}{\del r}=\frac 18 e^{4 A }\ 
\frac{1}{\r^4}\ \biggl(-\r^{12}\sinh^2(2\chi)\sin^2\theta +2\r^6 
\cosh(2\chi)(1+\sin^2\theta)+2\cos^2\theta\biggr) \,.
\end{split}
\eqlabel{wres1}
\end{equation}

\section{Conclusion}
In this paper we observed that certain 5d gauged supergravity 
flows on the background $R^{3,1}\times R_+$ can be deformed 
to flows on backgrounds $S^3\times R\times R_+$ or 
$dS_4\times R_+$ with the {\it same} 5d scalars. 
If the 10 dimensional lift of the  original backgrounds 
is known, this implies that deformed flows can be uplifted to ten 
dimensions as well. We explicitly demonstrated this for the 
$\caln=2^*$ PW flow, constructing for the first time 
massive RG flow with asymptotically global $AdS_5$ geometry. 
We hope that study of  these 
backgrounds would help develop gauge/gravity dictionary for 
gauge theories in curved space-time, including $dS_4$ deformations 
which might be relevant for understanding strings in backgrounds 
with cosmological horizons \cite{bds,blw}.

\section*{Acknowledgments}
It is a pleasure to thank Ofer Aharony for very interesting and stimulating 
discussions. I would like to thank the Weizmann Institute of Science, 
University of Pennsylvania for hospitality during part of this work.  
I would also like to thank the  Aspen Center 
for Physics  (2001 workshop) for hospitality, where this work started. 
I would like to thank Nick Evans for the e-mail correspondence concerning 
potential singularity of RG flow in \cite{evans}.

\section*{Appendix}
In \cite{evans}, Babington, Crooks and Evans
(BCE) proposed 
a supergravity dual to $\caln=0^*$ gauge theory. This gauge theory 
is obtained by giving the same mass to all four Weyl  fermions 
in the $\caln=4$ SYM theory. 
Note that this $\caln=4$ mass term completely 
breaks the supersymmetry,  hence the name for the deformation. 
In the infrared the $\caln=0^*$ gauge theory is expected to confine 
with  a mass gap in the spectrum. The mass gap in the gauge theory 
spectrum in particular  would imply that the dual, nonsupersymmetric,
supergravity background of \cite{evans} is nonetheless stable.
The $\caln=0^*$ dual supergravity solution is 
constructed first in 5d gauged supergravity and then was uplifted to
the full ten dimensional solution \cite{evans}. 
The authors turned on only the 5d scalar, called $\lambda$, dual to the 
$SO(4)$ invariant fermion mass term. They argued, based on the 
D3 brane probe computation, that the six $\caln=4$ scalars 
have positive radiatively induced (mass)$^2$, that thus, 
naively expected\footnote{
Note that this is precisely what is happening for the $\caln=2^*$
flow of \cite{pw}: the fermion mass term induces the {\it
negative} (mass)$^2$ term for scalars causing the D3 branes, originally 
at the origin, to 'spread' up in an enhancon configuration  
\eqref{rho}.}, expectation value for these scalars 
is not induced. In other words turning on $\lambda$ {\it alone} 
is consistent, and the confined $\caln=0^*$ vacuum is at the 
origin of the $\caln=4$ moduli space. 
   
If this would be the case, the supergravity solution 
\cite{evans} should have been nonsingular. 
We believe the solution of \cite{evans} has a naked singularity 
in the interior \cite{notes}. This in particular is 
reflected in the fact  that the dilaton of \cite{evans} is singular in 
the infrared, $|\lambda|\to \infty$,  for\footnote{For  notations 
see \cite{evans}.} $\alpha=\pi/2$.  The technical problem appears to 
be  attributed to the geodesic incompleteness of geometry \cite{evans}
for their choice of a radial coordinate, and thus incorrect identification 
of what is the infrared part of the geometry. From the physics
perspective, we suspect  that the probe computation in \cite{evans} 
is not a reliable tool to address the issue of radiatively induced 
scalar masses, due to the large nonperturbative corrections 
(fractional instantons) in theories with less then eight 
supercharges\footnote{This is briefly discussed in \cite{b}.}. 
In a sense solution \cite{evans} is akin to 
a singular Klebanov-Tseytlin (KT) geometry \cite{kt}, where unphysical 
requirement of unbroken chiral symmetry led to a singularity 
of the dual supergravity solution. In \cite{evans}, we believe this 
requirement is a zero expectation value for scalars. 
The analog of the Klebanov-Strassler (KS) \cite{ks}
resolution of the KT singularity in the BCE case is likely to 
be the inclusion of the additional 5d gauged supergravity 
scalar\footnote{Playing the role similar to $\rho$ 
for the $\caln=2^*$ PW flow \eqref{supp}.} 
corresponding to $SO(4)$ invariant 'distribution' 
of D3-branes. We expect the resulting nonsingular geometry 
will be similar to the polarized-branes solution of Polchinski 
and Strassler \cite{ps}.

While KT solution is unphysical in its original form, it is easy 
to turn it to a physical one by deforming the theory in such a way 
that the spontaneously broken chiral symmetry in the IR  gets 
restored. This can done by either considering sufficiently hot 
thermal state of the gauge theory \cite{b0011,ghkt}, or with  
an $S^3$ \cite{bt} or $dS_4$ \cite{bds} deformation of the 
gauge theory space-time. In the remaining of this section 
we argue that the prescription \cite{bt,bds} resolves 
BCE singularity. The latter is a reflection of the  
expectation that the infrared cutoff due to the $S^3$ scale 
or the Hubble parameter in the $dS_4$ deformation would stabilize 
the $\caln=4$ scalar masses for the $\caln=0^*$ flow,
resulting, as proposed in \cite{evans}, in the vacuum at the 
origin of the $\caln=4$ moduli space. 
We demonstrate this in the $S^3$  
deformed 5d gauged supergravity solution of 
BCE\footnote{There is a nonsingular lift of these  deformed 5d 
solutions.}.
 
The 5d solution of BCE comes from the effective action         
\begin{equation}
S=\int d\xi^5 \sqrt{-g}\left(\frac 14  R-\frac 12(\del\lambda)^2
-\calp\right)\,,
\eqlabel{action5e}
\end{equation} 
where the potential $\calp$ is
\begin{equation}
\calp=-\frac 32 \left(1+\cosh^2\lambda\right)\,,
\eqlabel{ppe}
\end{equation}
and  the flow metric 
\begin{equation}
ds_5^2=e^{2 A} \left(-dt^2 +d\bar{x}^2\right)+dr^2\,.
\eqlabel{flowmetrice}
\end{equation}
For the $ R^{3,1}\to R\times S^3$ deformation\footnote{The $\ R^{3,1}\to 
dS_4$ deformation can be analyzed in a similar fashion.} \eqref{ab}, 
 the equations of motion are 
\begin{equation}
\begin{split}
0&=\lambda''+\left(4 A'+3 B'\right) \lambda' -\frac{\del \calp}{\del
\lambda}\,,\\
0&=B''+4 A' B'+3\left(B'\right)^2-2 e^{-2 A -2 B}\,,\\
&\frac 14 A''+\left(A'\right)^2+\frac 34 A' B'=-\frac 13 \calp\,,\\
&- A''-\left(A'\right)^2-\frac 32 A' B'-\frac 34 B''-\frac 34 
\left(B'\right)^2=\frac 12 \left(\lambda'\right)^2 +\frac 13 \calp\,.
\end{split}
\eqlabel{ca} 
\end{equation}
Though we can't find an exact analytical solution of \eqref{ca}, 
we can still argue that the corresponding backgrounds are nonsingular,
provided we choose infrared boundary condition $B|_{r=0}=0$.
In the infrared ($r\to 0$) these nonsingular solutions form 
a one-parameter $\{\lambda_0\}$ family of  
Taylor series expansions\footnote{By rescaling the time coordinate 
in \eqref{ab} we 
can always set $A|_{r=0}=0$.}  
\begin{equation}
\begin{split}
e^A&=1+\left(\sum_{i=1}^{\infty}\ \a_i\ r^{2 i}\right)\,,\\
e^B&=r\left(1+\sum_{i=1}^{\infty}\ b_i\ r^{2 i} \right)\,,\\
\r&=\lambda_0+\left(\sum_{i=1}^{\infty}\ \lambda_i\ r^{2 i}\right)\,,\\
\end{split}
\eqlabel{ds0ase}
\end{equation}
with the first terms being 
\begin{equation}
\begin{split}
a_1&=\frac{3}{8}+\frac{1}{8}\  \cosh(2\lambda_0)\,,\\
b_1&=-\frac{1}{4}-\frac{1}{12}\  \cosh(2\lambda_0)\,,\\
\lambda_1&=-\frac{3}{16}\ \sinh(2\lambda_0)\,.\\
\end{split}
\eqlabel{1dse}
\end{equation}
Here $\lambda_0=0$, corresponds to global  $AdS_5$ solution.
Numerical analysis of \eqref{ca} with boundary condition 
\eqref{ds0ase} suggests that for arbitrary value of $\lambda_0$,
in the ultraviolet ($r\to \infty$) we obtain UV asymptotics of \cite{evans}
\begin{equation}
B\to 1,\ A\to \frac 12 r,\ \lambda\propto e^{-r}\,.
\eqlabel{UVe}
\end{equation}

\end{document}